# Dynamics of polar vortex crystallization

S. Rijal[1]†, Y. Nahas[1,2] †, S. Prokhorenko[1,2] *, and L. Bellaiche[1,2]

[1]Physics Department, University of Arkansas; Fayetteville, Arkansas 72701, United States

[2]Institute for Nanoscience and Engineering, University of Arkansas; Fayetteville, Arkansas 72701, United States

†These Authors contributed equally: S. Rijal, Y. Nahas

*Corresponding Author: S. Prokhorenko, Email: prokhorenko.s@gmail.com

## Abstract

Vortex crystals are commonly observed in ultra-thin ferroelectrics. However, a clear physical picture of origin of this topological state is currently lacking. Here, we show that vortex crystallization in ultra-thin $Pb(Zr_{0.4},Ti_{0.6})O_3$ films stems from the softening of a phonon mode and can be thus described as a SU(2) symmetry-breaking transition. This result sheds light on the topology of the polar vortex patterns and bridges polar vortices with smectic phases, spin spirals, and other modulated states. Finally, we predict an *ac*-field driven resonant switching of the vortex tube orientation which could enable new low-power electronic technologies.

Periodic stackings of polar vortex tubes were first predicted to form in atomically thin Pb(Zr,Ti)O$_3$ (PZT) films [1,2] and recently observed at room temperature in PbTiO$_3$/SrTiO$_3$ (PTO/STO) superlattices [3]. These emergent states, hereon referred to as vortex crystals (VC), can be controlled by several tuning factors [4–7] and are extensively researched for their technologically prominent properties, such as collective vortexon excitations [8], coupling to light [9] and electrons [10]. Interestingly, apart from perfectly periodic crystalline arrangements, vortex tubes were also shown to form glass- and liquid-like phases. Namely, Nahas *et al.* [11] have recently shown that out-of-equilibrium cooling of ultra-thin ferroelectric films can result in meandering vortex tube arrangements named polar labyrinths. These phases are characterized by a well-defined vortex tube structure at the local scale as well as a short-range ordering of vortex tubes but lack the long-range periodicity of vortex crystals. The same study has revealed glass-like dynamical properties of polar labyrinths – a low temperature ($T$) kinetic arrest and dynamical re-organization of the vortex tubes at finite temperatures was predicted and observed in ultra-thin PZT and BiFeO$_3$ films [11]. These finding allowed for the discovery of an inverse transition linking polar labyrinths and VCs. Finally, Zubko *et al* [12] have presented numerical evidence of thermally activated motion of polar vortices in PTO/STO superlattices reminiscent of a melting process. Such VC to a vortex liquid transition was further recently studied by Gómez-Ortiz *et al* [13].

The discoveries of polar vortex crystals, glasses and liquids pose an interesting question about the nature of the meso-scale ordering of polar vortices. For example, these states have been recently shown to share a profound connection with polar bubbles [4,6,7,11,12] which is in many ways similar to the relation between the wave-like spin states (e.g., conical or helical phases) and magnetic bubbles or skyrmions [13–17]. Moreover, the established link [4] between vortex freezing and phase separation kinetics clearly showed that the formation of vortex patterns is intimately related to symmetry breaking. If so, what are the symmetries involved? Do vortex tubes behave as particles or shall rather be seen as "condensed" waves? And, finally, can answering these questions allow for a better understanding of vortex crystals, liquids and labyrinths and lead to a prediction of new technologically prominent phenomena?

Here, we answer all these questions using effective Hamiltonian molecular dynamics simulations [2,14,15] of $PbZr_{0.4}Ti_{0.6}O_3$ (PZT) films. Our analysis of the finite temperature lattice excitations reveals that the melting of VC is associated with the softening of a phonon mode at a wave-vector away from the Brillouin zone center. Furthermore, the identified mode acquires a Goldstone character associated with an emergent continuous SU(2) symmetry at the transition temperature $T_C$. These results clarify the symmetry breaking mechanism and provide a new physical interpretation of the polar vortex states as waves akin to spin spirals, charge density waves and density modulations in liquid crystals. Finally, the proposed physical interpretation enables us to predict and numerically confirm a new phenomenon - the resonant switching of the vortex crystal orientation with low-amplitude *ac* fields.

First, we start by carefully examining the structure of polar vortex crystals. For this, we perform simulations of 5 unit cell (~2 nm) thick PZT films (see Methods in Supplementary Materials (SM) [16] ). Such films are assumed to be epitaxially grown under a misfit strain of -2% and have partially screened interfaces with a realistic 80% reduction of the surface bound charges. The supercell, initially thermalized at 800K, is cooled down to 25K at the steps of 25K. In accord with previous studies [2,3,17,18] on PZT and other ultrathin ferroelectrics, we find that the system undergoes a phase transition from a paraelectric-like state to a vortex crystal at $T_C \sim$ 369K.

In **Fig. 1a**, we show a typical calculated structure of such VC state at $T = 25K$. The corresponding polar pattern can be described as a periodic stacking of clockwise-anticlockwise vortex tubes formed by local electric dipoles. The axes of such tubes are parallel either to the *x* or *y* in-plane directions ($[100]_{p.c.}$ or $[010]_{p.c.}$ crystallographic axis, respectively) with the vortices forming in the (*x*, *z*) (respectively, (*y*, *z*)) cross sections of the film (**Fig. 1b**). Either of the two equivalent orientations of vortex tubes is arbitrarily chosen by the system at the transition. In this particular simulation, we obtain vortex tubes extending along the *y* axis and stacked in the *x* direction.

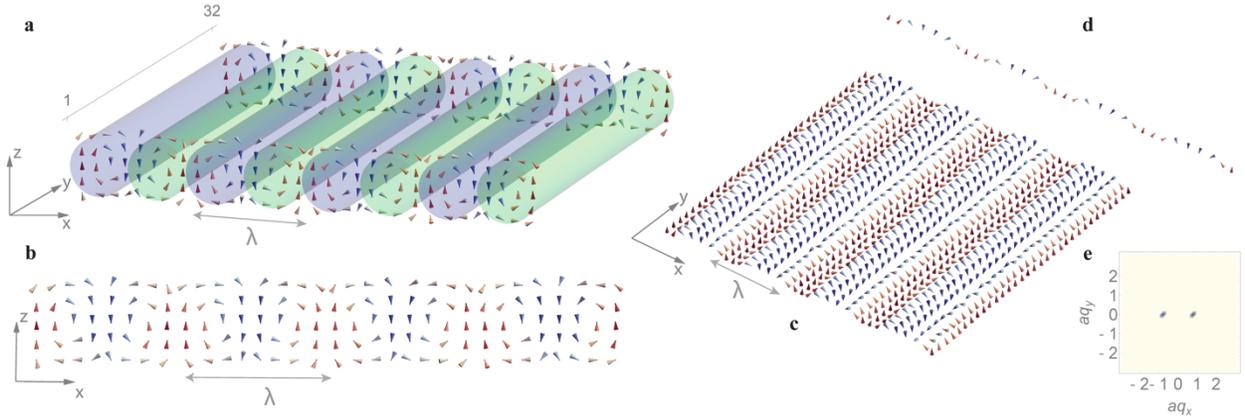

**FIG 1.** (a) VC state in a 32 x 32 x 5 supercell at 25K. The vortices made of local electric dipoles are seen on the xz faces (only y=1 and y=32 planes are shown here). Blue (Green) tube represents clockwise (anticlockwise) vortex tube with vortex axis parallel to y-axis; (b) and (c), respectively, show a xz (y=8) and xy (z=4) plane of the supercell of panel (a) where cycloidal periodic modulation of x and z components of dipoles is seen along the x-direction with a period of length $\lambda \sim 8a$. (d) shows the cycloidal modulation of dipoles along a straight line (line of intersection of z=4 and y=1 planes) along the x-direction. Colors of arrows in (a), (b), (c) and (d) represent direction of the z-component of the dipoles where shades of red (blue) denote +z (-z) direction. (e) The structure factor plot obtained by Fourier Transform of the z-component of the dipolar field in panel (c).

When looking at the ($x$, $y$) cross section of the film (e.g., **Fig. 1c**), one observes that the local dipoles form alternating stripes of "up" (+z, red) and "down" (-z, blue) pointing dipoles. Originating from the difference in the two perspectives, *i.e.*, that of vortices seen in the (*x,z*) plane and stripes seen in the (*x,y*) plane, the VC states have been referred to as both polar vortices [3] and stripe domains [1] in the literature. However, a close look also reveals that the electric dipoles in all (*x,y*) planes are harmonically modulated with a period of $\lambda \sim 8$ unit cells. For example, in **Fig. 1c**, one can clearly see a cycloidal wave in the $z = 4$ plane wherein the dipoles exhibit Néel rotations upon traversing the "stripes" as highlighted in **Fig. 1d**. This fact is also mirrored in the amplitude $S_q$ of the Fourier transform of the *z*-component of local dipoles shown in **Fig. 1e**. The only non-zero $S_q$ values are seen along the x-direction corresponding to a wavevector of magnitude $q_x = \frac{2\pi}{8a}$ which proves a harmonic wave character of the pattern. Hence, the polar structure in the $z = 4$ plane perfectly matches the distribution of spins in the so-called spin cycloids [19,20]. Interestingly, similar observations have been also made from STEM imaging of vortex crystals in $PbTiO_3/SrTiO_3$ superlattices [21]. However, in the latter case, the dipole modulations also possess an additional Bloch [22] component (rotation about the direction of modulation) due to which the wave pattern is rather helical [23,24]. At the same time, one important difference of the polar VC patterns from e.g. spin cycloids resides in the changing sense of Néel rotation within the bottom- and top halves of the film – the dipoles rotate clockwise (anti-clockwise) in the top (bottom) layers with increasing *x*.

Nonetheless, a mere observation of the structure begs clarification on whether the VC state emerges from a collective excitation of electric dipoles similar to how spin cycloids, helices and skyrmions found in magnetic counterparts result from collective excitations of their spins [25–28]. To gain further insight on this matter we thus inquire into the phonon excitation spectra of the VC state. For this, we compute the phonon spectral energy density [29,30] (SED) $\Phi$ for the simulated vortex crystal shown in **Fig. 1a**. $\Phi$ is defined as the average kinetic energy of electric dipoles per unit cell at a given wavevector **q** and frequency $\nu$. Here, we also separate the SED into contributions $\Phi^\alpha(\nu, \mathbf{q})$ from the vibration of the $\alpha = x, y, z$ components of the dipoles. Such component-based division of the spectral energy density allows us to assess the character of each mode. The component resolved SED calculated at 25K along the $q_x$ and $q_y$ directions is shown in **Figs. 2a** and **2b**, respectively, for $\alpha = $ x, y, z. The lowest frequency polar mode (marked by yellow circles) is revealed in the $\Phi^x$ and $\Phi^z$ plots along the $q_x$ direction at the wavevector $q_0 = \frac{2\pi}{8a}$ corresponding to the periodicity of the VC state. Moreover, the same mode is not visible in the $\Phi^y$ plot (**Fig. S1 in SM** [16]). This indicates that the dipoles fluctuate in the (x,z) plane, which is also the plane of dipolar rotations forming the vortex pattern. Hence, the lowest energy polar mode bears a striking similarity with the VC structure itself.

To corroborate this observation, we further extract the eigenvector associated with the described mode. Schematically shown in **Fig. 2c**, the eigenvector (red or blue arrows) is overlaid with the VC dipolar structure (gray arrows). The colored (gray) arrows thus represent the dynamic (static) component of the dipoles. As one can see, the lowest energy mode at the $q_x = q_0$ wavevector effectively amounts to the change of the dipole amplitude throughout the material. All of the dipoles first coherently increase in magnitude (red arrows) and then, half a period later, their magnitudes coherently decrease (blue arrows) with the frequency of 1.5 THz. Essentially, this means that the mode in question describes the oscillation of the amplitude of the VC modulation throughout the material. A recent study on a $PbTiO_3/SrTiO_3$ system [31] with polar vortex ground state also reports a similar eigenvector indicating a similarity in the VC states in PZT and PTO-based systems. Finally, we note the presence of a minimum in the $\Phi^y$ and $\Phi^z$ plots along the $q_y$ direction (marked by orange circles in **Figs. 2b**). As we will show below, the corresponding $q_y$ branch is, in fact, a relic of the degeneracy between the *x*- and *y*-oriented vortex tube crystals.

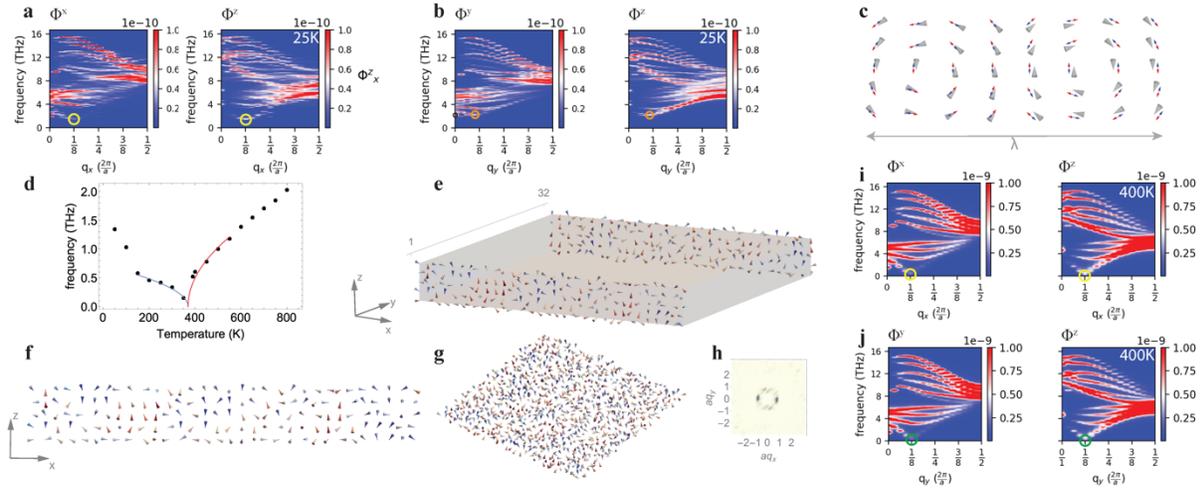

**FIG. 2.** (a-b) are spectral energy density plots of a 32 x 32 x 5 PZT supercell at 25K corresponding to a simulation time of 0.6 ns. Colors represent magnitude of SED of the $\alpha$-component (x, y or z) of the dipoles, $\Phi^\alpha(\nu, q_\beta)$, defined as a function of frequency, $\nu$, and wavevector, $q_\beta$, where $\beta$=x,y represents the wavevector direction. Yellow circles in (a) highlight the mode responsible for the VC formation. Black circle in $\Phi^y$ of (b) represents the frequency of externally applied homogeneous electric field. Orange circles in (b) represent a competing VC mode, (c) is the schematic diagram of a typical clockwise- anticlockwise dipolar vortex pair, as shown by gray cones, seen on the (x,z) [or (x,y)] cross-section of ultrathin PZT films. Upon each dipole (gray cone) is laid a double sided blue-red arrow that represents their periodic fluctuation in time where the blue and the red arrows are collective fluctuations separated by half a period. (d) Frequency versus temperature diagram of the soft mode responsible for VC; (e) 32 x 32 x 5 supercell at 400K, averaged over 60 ps. (f) and (g) show, respectively, a middle (x,z) and (x,y) cross-section of the supercell of panel (e). (h) is the Structure factor plot obtained by Fourier Transform of the z component of the dipolar field in panel (g). Colors in (e), (f) and (g) represent z-component of the dipoles with shades of red (blue) representing +z (-z) direction. (i-j) are spectral energy density plots of a 32 x 32 x 5 PZT supercell at 400K corresponding to a simulation time of 0.6 ns.

We now turn to the behavior of the frequency of the identified vortex mode (V mode) with temperature $T$. The data obtained from finite temperature MD simulations is shown in **Fig. 2d**. Upon increasing the temperature, the frequency of the V mode gradually decreases and approaches zero at $T\sim369K$. Such decrease upon approaching the critical temperature $T_c$ is typical of soft phonon modes responsible for second order phase transitions. Furthermore, and as shown by red (blue) fit lines in **Fig. 2d**, the softening follows the square root law $v\sim\sqrt{|T_c - T|}$ both below and above $T_c\sim369K$ which is also consistent with mean-field approximation of soft mode theory [32]. Interestingly, similar behavior was also previously reported for dynamical formation of vortex tubes in ferroelectric nanowires [33]. For the temperatures approaching $T_c$ from above, a clearly defined polar vortices in the (x,z) planes as well as wave patterns in the (x,y) planes are not observable. At the first sight, the equilibrium dipolar structure appears to be disordered (e.g., the dipole configuration averaged over 50ps at T=400K is shown in **Figs. 2f-g**). However, the calculated structure factor (**Fig. 2h**) possesses clear signatures of the periodicity of the VC ground state occurring below $T_c$. Specifically, $S_q$ acquires the shape of a ring with the radius $|q|\sim q_0 = \frac{2\pi}{8a}$. The latter observation points to a liquid-like phase consisting of a superposition of the competing long-range ordering modes, but that has not condensed yet along any particular direction.

**Figures 2i-j** contain the dispersion relations of optical phonons at $T = 400K$ and are representative of the temperature region within the upper vicinity of $T_c$. Beyond the observation of the softening of the V mode frequency in **Fig. 2d**, one can clearly see that the dispersion of the VC branch becomes quasi-linear in the vicinity of $q_x\sim q_0$ and $q_y\sim q_0$ (**Figs. 2i-j**). Such behavior is typical of gapless Goldstone modes responsible for the breaking of continuous symmetries, which, in our case consists of (*1*) the translational invariance of the liquid state at $T > T_c$ (2) rotational symmetry relating the vortex crystals with tube axis oriented either along the *x* or *y* axis.

Such symmetry explains why the dispersions along the $q_x$ (**Fig. 2i**) and $q_y$ (**Fig. 2j**) axes are identical up to 90° in-plane rotation of the local dipoles. Above the transition, the V modes with the propagation vector along either $[100]_{p.c.}$ and $[010]_{p.c.}$ are indistinguishable and soften upon decreasing the temperature. Then, at $T=T_c$, the system relaxes to either one of the vortex crystal states and, upon further cooling, the previously degenerate V modes split into the two low-frequency excitation branches with similar, yet distinct dispersions (see yellow and orange circles in **Figs. 2a-b**).

Having corroborated the hypothesis of the wave origin of the polar vortex patterns, we now turn to possible implications of this idea. The competition of two symmetry related V modes clarifies that the transition can be described by a 2D complex order parameter that we will denote as $\boldsymbol{Q} = (Q_x, Q_y)$ (see Methods in SM [16]) while the vortex crystallization breaks the SU(2) symmetry. As a by-product, such description demystifies the so-far elusive order parameter space of the system - similarly to e.g. the smectic phase of liquid crystals [34], the order parameter space has the topology of a Klein bottle which readily explains the existence of disclination-like topological defects in VC states. Moreover, within this picture, polar labyrinths can be now seen as coexisting $Q_x \neq 0$ and $Q_y \neq 0$ domains while the superposition of $Q_x$ and $Q_y$ waves suggests an explanation of the recently double-Q modulated states [35,36]. From a more practical standpoint, the revealed order parameter also hints to the possibility of switching from a $Q_x \neq 0$ to a $Q_y \neq 0$ state below $T_c$. If possible, such switching could allow to make an in-plane rotation of the polar vortex tubes in the entire material by 90° and could hardly be imaginable if the VC state was a collection of independent domains or vortex tubes.

Noting that the lowest frequency modes at $\boldsymbol{q} = 0$ have predominantly in-plane polarization (**Figs. 2a-b**), it is natural to assume that the VC patterns would be highly susceptible to in-plane oriented homogeneous electric fields. Hence, to probe the possibility of VC switching, we chose to apply such field oriented along the vortex tube axes ($y$ direction) of the form $E_y(t) = E_0 \sin(2\pi \nu t)$. Here, $E_0$ is the amplitude of the field while $\nu$ denotes its frequency. **Figure 3a** shows a schematic illustration of this computational experiment performed at $T = 25$K. A particularly remarkable result is obtained when $E_0$ reaches the threshold value of $50 \times 10^6 \, V/m$ and the frequency $\nu \sim 1.92$ THz coincides with one of the low frequency peaks in $\Phi^y(\nu, \mathbf{0})$ (black circle in **Fig. 2b**). Under such conditions, the vortex tubes axes first progressively switch from the $y$- to the $x$-direction, and, at $t > 29$ ps a dynamic vortexon-like [8] oscillation sets in (**Figs. 3b-d**) (**Video S1 in SM** [16]). During such oscillation, at first, the radii of the vortices reduce while the clockwise (blue) and the anticlockwise (green) vortex tubes move in the opposite direction along the $z$-axis (**Fig. 3b**). Such displacements give rise to a dipolar wave, that meanders around the vortex tubes and yields a non-zero $y$-component of electric polarization $P_y$. Interestingly, a static state consisting of such meandering dipolar waves have also been reported [10,23,33,37] in different ferroelectric systems. When the field magnitude starts to decrease, the evolution of dipoles reverses and, a quarter of a period later, the system returns to the VC state with tubes extending along the $x$-axis (**Fig. 3c**). The evolution of the system during the second half period is symmetry equivalent. During this stage, the zig-zag vortex tube displacements are inverted - the clockwise (anti-clockwise) vortex tubes displace upwards (downwards) as shown in (**Fig. 3d**). Here, the direction of the dipolar wave is also reversed resulting in a negative polarization $P_y$. Finally, upon turning the electric field off**,** the dipolar pattern settles to the VC state with the vortex axes lying parallel to the $x$-axis which accomplishes the switching (**Fig. 3e**). The time evolution of $P_y$ overlayed with the applied field $E_y(t)$ is shown in **Fig. 3f**.

Furthermore, the evolution of the order parameter components $|Q_x|(t)$ and $|Q_y|(t)$ is shown in **Fig. 3g**. These plots clearly demonstrate that, in response to $E_y$, the initial $|Q_x| \neq 0$ state (red curve) vanishes and a new $|Q_y| \neq 0$ state emerges within the time interval 21 ps $< t <$ 29 ps. Enroute to switching the axis of the vortex tubes (see **Video S2 in SM** [16]), the system passes through some dynamical and topologically non-trivial transient states. **Figures 3(h-k)** show several of such dipolar structures including transient vortex tube deformations **(Fig. 3h)** and meron crystals **(Fig. 3i)** akin to the patterns found in PTO/STO systems [35,36] as well as polar labyrinths [4,11] **(Fig. 3j-k)** featuring polar disclinations [37] and bubble-like [4] geometries. A detailed analysis of these states opens additional avenues for future studies on these systems.

Notably, the switching between equivalent VC states is not observed for applied field frequencies that deviate from the frequency $\nu \sim 1.92$ THz. Also, the threshold $E_0$ value of $50 \times 10^6$ $V/m$ is an order of magnitude smaller than the critical out-of-plane fields triggering transitions to the bubble and monodomain states in this system. In this sense, the described switching appears to be a resonant phenomenon. Owing to highly anisotropic properties of vortex crystals, the revealed VC reorientation could lead to the design of low-power and ultrafast switches and signal processing devices.

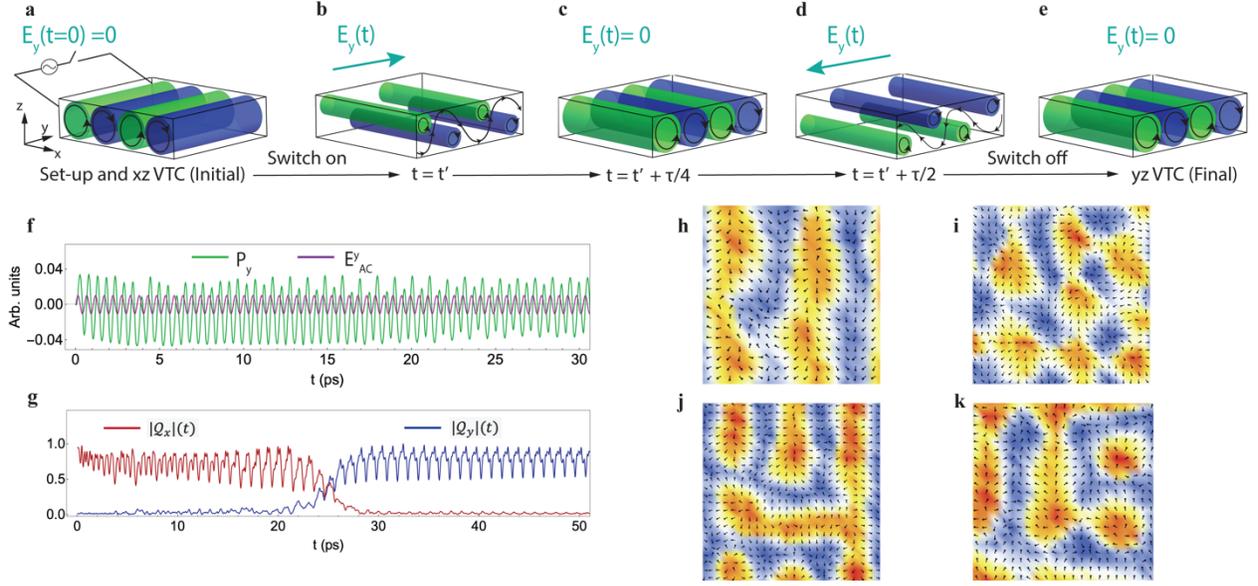

**FIG 3.** Schematic diagram (a) shows a cartoon set-up of application of time varying homogenous electric field, $E_y(t)$, at 25K, along the y direction. The initial VC state before the electric field is applied is such that the vortices are seen in the (x,z) plane and vortex tube axes are aligned along the y-direction; (b), (c) and (d), respectively, depict intermediate states when the magnitude of electric field is positive, zero and negative along the y direction, at times $t = t'$, $t = t' + \frac{\tau}{4}$ and $t = t' + \frac{\tau}{2}$, where $\tau$ is the period of the applied electric field. (e) shows the final VC state after the electric field is switched off, with vortices (vortex tube axes) having now switched from the (x,z) plane (y-direction) to the (y,z) plane (x-direction). (f) shows the y-component of polarization response (green) to the electric field (magenta). Note that the electric field is scaled such that the polarization response is clearly visible. (g) Time evolution of the order parameter components $|Q_x|(t)$ and $|Q_y|(t)$ of the VC states. (h-k) Show snapshots of a middle (x,y) plane during 10ps < t < 30 ps when the VC orientation is being switched by 90°. Positive (negative) z-component directions of the local dipoles represented by red (blue) color.

In summary, our results show that vortex crystals in ultra-thin ferroelectrics result from the condensation of soft optical phonons and can be perceived as waves rather than ensembles of domains or standalone vortex tubes. This finding establishes a link between topological states in ultra-thin ferroelectrics with their magnetic counterparts, but also opens new analogies with smectic phases of liquid crystals, charge density waves and other harmonically modulated states [38,39]. The proposed change of perspective can push both the theoretical and experimental researches on polar topologies along novel inter-disciplinary directions. Finally, our prediction of the ultrafast resonant switching of the vortex crystal orientation opens conceptually new possibilities for phonon engineering and functionalization of ferroelectrics in information storage and processing devices.

We acknowledge the Vannevar Bush Faculty Fellowship (VBFF) Grant No. N00014-20-1-2834 from the Department of Defense and an Impact Grant from Arkansas Research Alliance. We also appreciate the Arkansas High Performance Computing Center where the computations were performed.